\documentstyle{l-aa}
\begin{document}
\newcommand{\be}{\begin{eqnarray}}
\newcommand{\ee}{\end{eqnarray}}
\def\a{\"a}
\def\A{\"A}
\def\o{\"o}
\def\O{\"O}
\def\Omm{\Omega_m}
\def\Oml{\Omega_{\Lambda}}
\def\Om0{\Omega_0}
\def\Omb{\Omega_b}
\thesaurus{01(12.03.3; 03.13.6)}

\title{Statistical evaluation of the observational information on
$\Omm$ and $\Oml$}

\author{S. M. Harun-or-Rashid
\and M. Roos
}

\institute{Physics Department, High Energy Physics Division,
P.O.B. 64, FIN--00014 UNIVERSITY OF HELSINKI, Finland}

\offprints{S. M. Harun-or-Rashid, e-mail:
hsrashid@pcu.helsinki.fi}

\date{Received 06.02.2001 ; accepted 09.03.2001}

\maketitle

\begin{abstract}

We undertake a critical evaluation of recent observational
information on $\Omm$ and $\Oml$ in order to identify possible
sources of systematic errors and effects of simplified
statistical analyses. We combine observations for which the
results have been published in the form of likelihood contours in
the $\Omm ,\Oml$ plane. We approximate the contours by fifth order
polynomials, and we then use the maximum likelihood method to
obtain joint likelihood contours for the combined data. In the
choice of statistical merits we aim at minimum loss of information
rather than at minimum variance. We find that
$\Om0 = \Omm + \Oml = 0.99\pm 0.04 \pm 0.03$, where the first error
is mainly statistical and the second error is systematical. In a flat
Universe we find $\Omm^{flat} = 0.31 \pm 0.04 \pm 0.04$.


\end{abstract}

\section{INTRODUCTION}

Our knowledge of the dynamical parameters of the Universe
describing the cosmic expansion has improved rapidly over the
last few years, starting with the epochal discovery of the large
scale anisotropies of the CMB by COBE-DMR (Smoot et al. 1992),
followed by the dramatic supernova Ia observations by the High-z
Supernova Search Team (Riess et al. 1998) and the Supernova
Cosmology Project (Perlmutter et al. 1998, 1999), and most
recently by the measurements of the first acoustic peak in the
CMB power spectrum in the first results from the balloon flights
BOOMERANG (de Bernardis et al. 2000) and MAXIMA (Balbi et al.
2000, Hanany et al. 2000).

The list of other recent observations is very long, even if one
restricts oneself to those having information on both the mass
density parameter $\Omm$ and the density parameter $\Oml$ of
vacuum energy. Recall that $\Oml$ is related to the cosmological
constant $\Lambda$ by
\begin{eqnarray}
\Oml = \Lambda/3H^2_0\ . \label{f0}\end{eqnarray}
 Lineweaver (1998) and Tegmark (1999) have summarized and
analyzed some 20 more observations of the CMB anisotropies (cf.
 their reference lists). Determinations of $\Omm$ and $\Oml$
 have been reported from observations on the gas fraction in
 X-ray clusters (Evrard 1997), on X-ray cluster evolution
 (Bahcall \& Fan 1998, Eke et al. 1998), on the cluster
 mass function and the Ly$\alpha$ forest (Weinberg et al.
 1998), on gravitational lensing (Chiba \&
 Yoshi 1998, Helbig 2000, Im et al. 1997), on the Sunyaev-Zel'dovich
 effect (Birkinshaw 1999, Carlstrom et al. 1999), on classical
 double radio sources (Guerra et al. 2000), on galaxy
 peculiar velocities (Zehavi \& Dekel 1999), on the
 evolution of galaxies and star creation versus the evolution
 of galaxy luminosity densities (Totani 1997). The large
 scale structure and its power spectrum has been studied
 in the SSRS2 and CfA2 galaxy surveys (da Costa et al., 1994),
 in the Las Campanas Redshift Survey (Schectman et al., 1996),
 in the Abell-ACO cluster survey (Retzlaff et al., 1998), and
 in the 2dF Galaxy Redshift Survey (Peacock et al., 2001).

Some of the above information has already been used to
constrain $\Omm$ and $\Oml$, some of it could in principle
 be used that way, but has not been presented in a form readily
 useful to an analyst outside the observer teams. It must also
 be said that much is statistically weak, the analyses being
 simplified and the discussions of possible systematic errors
 absent.

Nevertheless, the list of large combined data analyses
since 1999 is already long. Lineweaver (1999) combined the SN Ia
data with CMB data, X-ray cluster data, cluster
evolution data and double radio sources. Le Dour et al. (2000)
analyzed only CMB data, whereas Tegmark et al.
(2000a) combined CMB data with IRAS LSS data. Tegmark \& Zaldarriaga
(2000b, 2000c) and Hu et al. (2000) combined BOOMERANG and MAXIMA
data, Melchiorri et al. (2000) combined BOOMERANG and COBE data.
Bridle et al. (2000) combined the CMB data with galaxy peculiar
velocities and with the SN Ia data. The BOOMERANG, MAXIMA
and COBE data have been combined with LSS and SN Ia data by
Jaffe et al. (2000) and Bond et al. (2000), and combined with
a different set of LSS data by Novosyadlyj et al. (2000a) and
Durrer \& Novosyadlyj (2000). In
a sequence of papers we (Roos \& Harun-or-Rashid 1998, 1999,
2000) combined much of the data quoted above having published
an error on $\Omm$ and $\Oml$, using simple $\chi^2$ analysis.
This of course implied believing in the errors and treating them
as Gaussian.

The conclusion of all these partially overlapping analyses is
that the Universe is consistent with being flat, $\Om0 = \Omm +
\Oml$ is near unity, and $\Oml/\Omm$ is near 2. But the analyses
differ in methods, in the treatment of errors and in their
attention to possible systematic errors, so the results differ in
the precision of these conclusions.

We undertake here yet another combined study using the maximum
likelihood method, and paying special attention to statistical
arguments. In Sec. 2 we discuss statistical methods in general
and present our method of analysis. In Sec. 3 we discuss the data
chosen for our analysis, all of which have been published
graphically as likelihood contours in the $\Omm , \Oml$ plane. In
Sec. 4 we give our estimates for the parameters, and in Sec. 5 we
discuss some related analyses.

\section{STATISTICAL METHODS}

The observational data generally contain information on a large
set of parameters. The information about each single parameter is
then obtained by fixing some parameters at known input values,
and marginalizing over others. Note that it is quite misleading
to report the values of each parameter in turn, always 
carrying out unconditional marginalizing over all the others,
because then the same information has been reused many times.
Already when marginalizing to obtain the value of the first
parameter one has used all the information available.

There is only one remedy to this: if one is mainly interested in
the values of a small subset of parameters, in our case two, one
should describe their joint pdf by confidence contours (or
surfaces or hypersurfaces) in the space of those parameters,
marginalizing over less interesting ancillary parameters. The
confidence range of the second parameter is then conditional on
the range of the first parameter, and so on. In practice all the
conditional confidence ranges are determined by the size of the
orthogonal box circumscribed around the two- (or higher-)
dimensional confidence contour.

In the present analysis we are only interested in the
values of $\Omm$ and $\Oml$, therefore we only use data for which
the marginalization over ancillary parameters has already been
carried out. Note that thereby we do use the full information of
each observation.

In several observations it has been noted that some parameters are
strongly correlated. If one marginalizes over one of a pair of
correlated parameters, the likelihood function of the other one
becomes quite broad. This is an effect we do not try to avoid,
because it implies including one type of systematic error.

Let us make a few comments of statistical nature comparing the
least squares or $\chi^2$ method with the maximum likelihood
method. The advantage of the least squares method is its
simplicity, it is unbiased, and a goodness-of-fit value can be
obtained by comparing the least squares sum with the number of
degrees of freedom. The disadvantage is that it requires the pdf
of the input data to be symmetric, preferably normal, but even
then there is no guarantee that the final estimate will be
normally distributed, except asymptotically. To form the least
squares of very conflicting data is statistically meaningless.

In contrast, the log-likelihood functions of any data can be
added, and if there are conflicts, due for instance to systematic
errors, they will show up as several dips. No goodness-of-fit
value can be obtained thereby one does not risk
statistically meaningless statements), but relative confidence
levels can be defined. There are no restrictions to the symmetry
or normality of the pdf of the input data. Asymptotically the
maximum likelihood estimator attains normality faster than the
least squares estimator.

When one compromises between different statistical merits, one
may conclude that it is more desirable to achieve minimum loss of
information than minimum variance. In the first case one wants to
make sure that no other single number could contain more
information about the parameter of interest than the estimate
chosen. In the second case one feels that the smaller the
variance, the more certain one is that the estimate is near the
true value. If one opts for minimum loss of information, the
maximum likelihood estimator is optimal in the asymptotic limit.

A problem that occurs in some of the data we use is that the pdf
extends into an unphysical region, specifically the region $\Omm
< 0$. Even the region $\Oml < 0$ might be considered unphysical in
this context. To simply ignore the unphysical region biases the
pdf and produces a systematic error. The remedy to this is the
method proposed by Feldman \& Cousins (1998). We, however, cannot
apply their method to the data we use, it has to be done at the
time of original data analysis. We shall only give mention where
it has been done, and where it should have been done.

Since none of the likelihood contours in the data we use look
normal nor even symmetric, we clearly choose the maximum
likelihood method. (The tool to use is actually not the
likelihood function which is the product of individual pdf's,
but the negative of the sum of their logarithms, or the
log-likelihood function.) We approximate the contours by general
fifth order polynomials of the form
\begin{eqnarray}
P(\Omm \Oml ) = \Omm^m \Oml^n ,\ \ \ m+n \le 5 .
\label{f1}\end{eqnarray} There are then 20 terms in the
polynomials, so we read off 20 points from the $1\sigma,\
1.64\sigma,\ 2\sigma,\ 3\sigma$ contours and the best value,
where available. Since we already know the approximate location
of the globally favored region from all the previous studies, it
is enough that we require our polynomial approximation to be good
over that region. This fit region is defined by $0.15 \le \Omm \le
0.50$ and $0.40 \le \Oml \le 0.88$, but the sample points are of
course taken also from outside this region in order to obtain a
well-behaved polynomial inside the region. Far away from it the
polynomial approximation of course breaks down completely.

One should recognize that the observational likelihood surfaces
in the $\Omm , \Oml$ plane are not known with a very good
resolution. Thus one cannot set very high requirements on the
polynomial representation inside the $1\sigma$ contours. We have
been checking that the polynomial is non-negative in the fit
region, and that its minimum (i.e. of the negative log-likelihood
function) is indeed at the observational best
value, where reported. We have also been checking the location of
the $0.33\sigma$ contour of the polynomial approximation, in
order to verify that it is reasonably centrally located with
respect to the $1\sigma$ contour and to the best value, when
reported.

\section{DATA}

Altogether we use six independent data sets meeting our criteria,
grouped into SN Ia data, CMB data, LSS data and other data. But
as we shall see, some of these data sets actually comprise
several other important observations.

\subsection{SUPERNOVA Ia DATA}

The SN Ia observations by the High-z Supernova Search Team (HSST)
of Riess et al. (1998) and the Supernova Cosmology Project (SCP)
of Perlmutter et al. (1998, 1999) are well enough known not to
require a detailed presentation here. The importance of these
observations lies in that they determine approximately the linear
combination $\Oml - \Omm$ which is orthogonal to $\Om0 = \Omm +
\Oml$.

HSST use two quite distinct methods of light-curve fitting to
determine the distance moduli of their 16 SNe Ia under study.
 Their luminosity distances are used to place constraints on
six cosmological parameters: $h, \Omm, \Oml, q_0,$ and the
dynamical age of the Universe, $t_0$.  The MLCS method involves
statistical methods at a more refined level than the more
empirical template model. The moduli are found from a $\chi^2$
analysis using an empirical model containing four free
parameters. The MLCS method and the template method give moduli
which differ by about $1\sigma$. Once the distance moduli are
known, the parameters h, $\Omm,\ \Oml$ are determined by a
maximum likelihood fit, and finally the Hubble constant is
integrated out. (The results are really independent of h.)
One may perhaps be somewhat concerned about the assumption that
each distance modulus is normally distributed. We have no reason
to doubt that, but if the iterative $\chi^2$ analysis has yielded
systematically skewed pdf's, then the maximum likelihood fit will
amplify the skewness.

The authors state that "the dominant source of statistical
uncertainty is the extinction measurement". The main doubt raised
about the SN Ia observations is the risk that (part of) the
reddening of the SNe Ia could be caused by intervening dust
rather than by the cosmological expansion. Among the possible
systematic errors investigated is also that associated with
extinction. No systematic error is found to be important
 here, but for such a small sample of SNe Ia one can
expect that the selection bias might be the largest problem.

The authors do not express any view about which method should be
considered more reliable, thus noting that "we must consider the
difference between the cosmological constraints reached from the
two fitting methods to be a systematic uncertainty". We shall
come back to this question later. Here we would like to point out
that if one corrects for the unphysical region $\Omm < 0$ using
the method of Feldman \& Cousins (1998), the best value and the
confidence contours will be shifted slightly towards higher
values of $\Omm$. This shift will be more important for the MLCS
method than for the template method, because the former extends
deeper into the unphysical region.

The likelihood contours in the $\Omm, \Oml$ plane (Riess et
al. 1998, Fig. 6) correspond to 68.3\%, 95.4\% and 99.7\%
confidence, respectively ($1\sigma$, $2\sigma$, $3\sigma$).

Let us now turn to SCP, which studied 42 SNe Ia. The MLCS method
described above is basically repeated, but modified in many
details for which we refer the reader to the source. The distance
moduli are again found from a $\chi^2$ analysis using an
empirical model containing four free parameters, but this model
is slightly different from the HSST treatment. The parameters
$\Omm$ and $\Oml$ are then determined by a maximum likelihood fit
to four parameters, of which the parameters $\mathcal{M}_B$
 (an absolute magnitude) and $\alpha$ (the slope of the
width-luminosity relation) are just ancillary variables which
are integrated out (h does not enter at all). The authors then
correct the resulting likelihood contours for the unphysical
region $\Omm < 0$ using the method of Feldman \& Cousins (1998).
The likelihood contours in the $\Omm, \Oml$ plane
(Perlmutter et al. 1999, Fig. 7) correspond to 68\% ($1\sigma$),
90\% ($1.64\sigma$), 95\% ($1.96\sigma$), and 99\% ($2.58\sigma$)
confidence, respectively.

Since the number of SNe Ia is here so much larger than in HSST,
the effects of selection and of possible systematic errors can be
investigated more thoroughly. SCP quotes a total possible systematic
uncertainty to $\Omm^{flat}$ and $\Oml^{flat}$ of 0.05.

If we compare the observations along the line defining a flat
Universe, SCP finds $\Oml - \Omm = 0.44 \pm 0.085 \pm 0.05$,
whereas HSST finds $\Oml - \Omm = 0.36 \pm 0.10$ for the MLCS
method and $\Oml - \Omm = 0.68 \pm 0.09$ for the template method.
This comparison tells us that the template method is afflicted by
systematic errors of its own. We choose the former since this
method is basically the same as used by SCP. This is admittedly a 
selection bias of
ours, but we shall account for it, at least partly, by applying
the same systematic error of $\Delta \Omm^{flat} = \Delta
\Oml^{flat}=0.05 $ to HSST as to SCP.

SCP and HSST then agree within their statistical errors -- how well
they agree cannot be established since they are not
completely independent. Part of the difference may be explained
by selection bias in the smaller set of SNe Ia.

\begin{figure}[h]
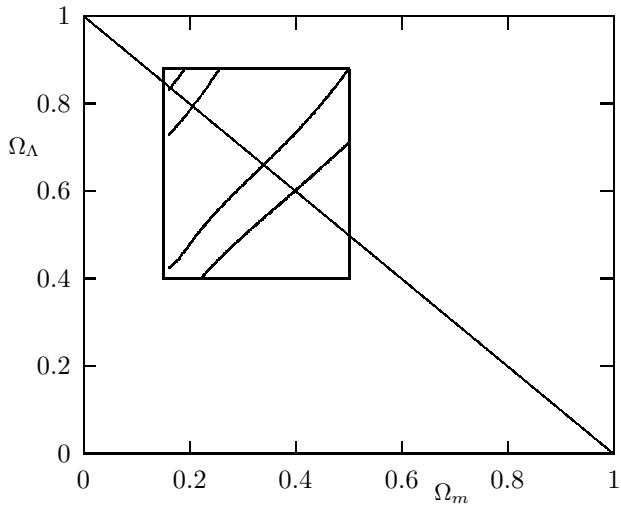

\input SNsig12.tex
\caption{The confidence contours of the log-likelihood sum of the
two SN Ia observations (HSST and SCP). The curves correspond to
1$\sigma$ and 2$\sigma$ in the ($\Omega_m,
\Omega_{\Lambda}$)-plane. The significance of the square is
described in the text. The diagonal line corresponds to a flat
cosmology.} \label{fig-1}
\end{figure}

In Fig. \ref{fig-1} we show the confidence contours of the log-likelihood
sum of the two SN Ia observations in our polynomial approximation,
drawn only in the ranges of $\Omm$ and $\Oml$ that we sample.
Along the flat line these observations determine $\Oml - \Omm =
0.45 \pm 0.13$. Note that our value is not obtained from the
weighted mean of the SCP and HSST values, but from our
log-likelihood sum.

There are many other types of observations which give
complementary information in support of the SN Ia data. These
observations have been summarized briefly by Perlmutter et al.
(1998, 1999), and in more detail for instance by us (Roos \&
Harun-or-Rashid 1998, 1999, 2000). But there have also been
gravitational lensing data in strong conflict with the SN Ia
data. The best value in our Fig. \ref{fig-1} is excluded with 99.7\%
confidence by the joint optical (spiral galaxies) and radio data
of Falco, Kochanek \& Muñoz (1998)(six gravitational lenses
analyzed). However, the authors point out that these results
depend on the choice of galaxy sub-type luminosity functions in
the lens models. Subsequently Chiba \& Yoshii (1999) have
emphasized this point in an analysis with E/S0 luminosity
functions that yielded a best fit mass density in a flat
cosmology, finding $\Oml - \Omm = 0.4 + 0.2/-0.4$ in agreement
with the SN Ia data.

More recently Helbig (2000) has shown preliminary results from the
Cosmic Lens All-Sky Survey (CLASS) of radio lenses only. These
results are still inaccurate, $-0.8 < \Oml - \Omm < 0.3$ at 95\% confidence,
but appear to be in strong conflict with the SN Ia data. It
is still too early to say whether SNe Ia will have to come down to smaller
values of $\Oml - \Omm$, or whether gravitational lensing will
have to go up. In Section 5 we shall discuss how our combined
fit changes when including the preliminary CLASS constraints.

\subsection{CMB DATA}

Before the advent of the balloon observations BOOMER\-ANG (de Bernardis et 
al. 2000) and MAXIMA-1 (Balbi et al. 2000, Hanany etal. 2000), 
Lineweaver (1998) and Tegmark (1999) analyzed all the
then existing CMB data in the form of multipole power spectra up
to $\ell \simeq 800$. The parameter space is then very
large, but some parameters really drop out and others can be
handled in a simplified manner if their effect is significant
only below or above $\ell \simeq 100$. In the analysis of
Tegmark (1999), the following ten cosmological parameters are
jointly constrained: $\Omega_k, \Oml,$ the optical depth parameter
$\tau,$ the amplitudes and slopes $A_s, n_s, A_t, n_t$ of scalar
and tensor fluctuations, and the physical matter densities
$\omega_b, \omega_{cdm}, \omega_{\nu}$. Of these parameters only
six are well constrained; the resulting 6-dimensional likelihood
function is then integrated over remaining parameters, and it is
stated to be highly non-Gaussian in some directions. We trust
that the plotted marginalized 2-dimensional confidence limits in
the $\Omm, \Oml$ plane are then realistic, and do not contain any
imposed Gaussian form.

The approximations made are claimed to reproduce the power
spectrum to about 5\% accuracy. Otherwise no systematic errors
are discussed. But the input data show rather large scatter, so
one might hope that by combining them, most of the systematic
differences between them have been taken into account. One worry
is that the best fitting models all fail to quite match the COBE
DMR data (Tegmark 1999).

The likelihood contours in the $\Omm, \Oml$ plane from this
compilation (Tegmark 1999, Fig. 3) correspond to 68\%
($1\sigma$) and 95\% ($1.96\sigma$) confidence, respectively. No
best value point is given. 

The balloon observations BOOMERANG (de Bernardis et al. 2000) and
MAXIMA-1 (Balbi et al. 2000, Hanany et al. 2000) have produced
the first high-resolution, high signal-to-noise maps of the CMB
from independent patches of the sky, and thereby derived the
angular power spectrum with a better precision than was
achievable in compilations of earlier observations. In both cases
the so far published results represent a complete analysis of a
limited portion of the data.

Given the multipole spectrum, Balbi et al. (2000) fit
different 7-dimensional CDM models to some pixelization of the
measured range of \textit{l}, including the COBE DMR data.
The jointly constrained cosmological parameters are $\tau, \Omb,
\Omm, \Oml, n_s$ and $C_{10}$, the amplitude of fluctuations at
multipole $\ell=10$. For a seventh parameter they alternatively
used $h$ and $\Omb h^2$. Marginalizing over five parameters, the
 confidence range in the $\Omm, \Oml$ plane were found. The
first acoustic peak is the dominant feature in the power
spectrum, the maximum occurring at $\ell = 197 \pm 6$ in
BOOMERANG and at $\ell \simeq 220$ in MAXIMA-1. The position of
the peak determines $\Om0$ pretty independently of all other
parameters; thus for a flat universe it determines $\Omm$.

BOOMERANG (de Bernardis et al. 2000) jointly constrain a
six-dimensional parameter space, comprising $h, \Omb h^2, \Omm,
\Oml, n_s$ and an overall normalization $A$. Subsequently they
marginalize over four parameters to obtain the confidence range
in the $\Omm, \Oml$ plane plotted in their Fig. 3.

The difference in $\Om0$ between the two BOOMERANG and
MAXIMA-1 is rather large, since $\ell \simeq 200 \Om0^{-1.58}$
for $\Omm = 0.3$ and $\Om0$ near 1 (Weinberg 2000). (In the
literature one sometimes sees the relation $\ell \simeq 200
\Om0^{-0.5}$ which is true only when $\Oml = 0$.) This difference
clearly represents a systematic error which should be allowed to
affect the total fit. All other systematic errors discussed have
less influence on $\Omm$ and $\Om0$. The likelihood contours in
the $\Omm, \Oml$ plane from MAXIMA-1 (Balbi et al. 2000)
correspond to 68\% ($1\sigma$), 95\% ($1.96\sigma$), and 99\%
($2.58\sigma$) confidence, respectively, which we can use. No
best value point is given.

BOOMERANG has only published a coarsely pixelized likelihood
surface of 95\% ($1.96\sigma$) confidence (de Bernardis et
al. 2000, Fig. 3), which would make our polynomial fit a poor
approximation. Therefore we do not include BOOMERANG as an
independent constraint. However, it is included together with the
LSS constraint to be discussed in the next subsection, so the
BOOMERANG information is not neglected.

A totally different approach is taken by Jaffe et al. (2000). The
two data sets both have a calibration uncertainty, 20\% for
BOOMERANG and 8\% for MAXIMA-1, which Jaffe et al. (2000) uses to
adjust the power spectra so that the peaks are more similar in
amplitude. The data can then be combined into multipole bands, and
the goodness-of-fit improves considerably. However, since the
relative importance of the two peaks is then altered, the
combined data yields a significantly shifted $\Omm$, and the
originally visible systematic difference in $\Omm$ disappears. We
fear that this leads to an underestimation of the $\Omm$
systematic error. As stressed before, we prefer to exhibit all
systematic errors in order to achieve minimum loss of information.

\subsection{LSS AND OTHER DATA}

An important source of information on $\Omm$ and $\Om0$ is the
power spectrum of matter density fluctuations. Durrer \&
Novosyadlyj (1999) have chosen to study this for Abell-ACO
clusters (Retzlaff et al., 1998) , arguing that the power
spectrum of clusters should better represent the observed
Universe as a whole, rather than density fluctuations on the
scale of galaxies. Making this choice may represent a bias
implying ignoring some inherent systematic error, but this seems
the only choice for us at the moment.

Novosyadlyj et al. (2000a) combine the power spectrum of Abell-ACO
clusters with six independent constraints for the amplitude of
the fluctuation power spectrum on different scales: from clusters
at different redshifts, from quasar spectra, and from the bulk
flow of galaxies in our vicinity (cf. their reference list). Very
importantly, they also use
the CMB value  $\ell = 197 \pm 6$ for the multipole moment of the
first acoustic peak from BOOMERANG which thus gets included into
our data base. In addition they constrain the Hubble constant to
be $h = 0.65 \pm 0.10$, a low but reasonable compromise value
with a generous error. They constrain the baryon density to be
$\Omega_b h^2 = 0.020 \pm 0.002 \ (95\% CL)$ (Burles et al. 2001).

The MAXIMA-1 and BOOMERANG values for $\Omega_b h^2$ are in stark
contradiction to that value, but this conflict does not affect
the value of $\Om0$ noticeably. In fact, the outcome of these
authors' multiparameter least squares fit to the cluster
power spectrum yields a baryon density more similar to the
MAXIMA-1 and BOOMERANG value, and yields a higher Hubble constant
closer to present best determinations.

It is not clear whether any systematic errors are included by
Novosyadlyj et al. (2000a). However, the fit errors are rather
large anyway, e.g. $\Omm = 0.37^{+0.25}_{-0.15}$ marginalized
over all other parameters, so that the influence of systematic
errors would be small. The information  in Novosyadlyj et al.
(2000b), Fig. 1a are the $1\sigma, 2\sigma$ and $3\sigma$
confidence contours in the $\Omm, \Oml$ plane, obtained by
marginalization over the other five parameters (and assuming that
only one species of massive neutrinos contribute to the neutrino
density parameter $\Omega_{\nu}$).

Our last constraint is based on the measurements of the
coordinate distance to sources at redshifts between 1 and 2,
which depends on the global values of the cosmological
parameters, but which is independent of the power spectrum of
density fluctuations, and of whether matter is biased relative to
light. From a parent population of 70 powerful extended classical
double radio galaxies Guerra et al. (2000) have studied a subset
of 20, for which it was possible to estimate independently the
mean and characteristic size, and thus the coordinate distance.
There is one model parameter $\beta$ in the theory in addition to
$\Omm$ and $\Oml$. The main systematic error is the model
uncertainty in $\beta$, but it is shown to be unimportant
compared to known statistical errors, and quite uncorrelated to
$\Omm$ and $\Oml$.

The information in Guerra et al. (2000), Fig 11, are the
68\% ($1\sigma$) and 90\% ($1.64\sigma$) likelihood contours in
the $\Omm , \Oml$ plane, obtained by marginalizing over $\beta$.
The favored region is quite large, so that this constraint is at
present quite weak, adding only some preference for small $\Omm$
values. A best value is claimed at $\Omm = 0$ and $\Oml = 0.45$.
It is not clear to us how the unphysical region $\Omm < 0$ has
been treated, in any case not with the Feldman -- Cousins (1998)
procedure.

\begin{figure}[h]
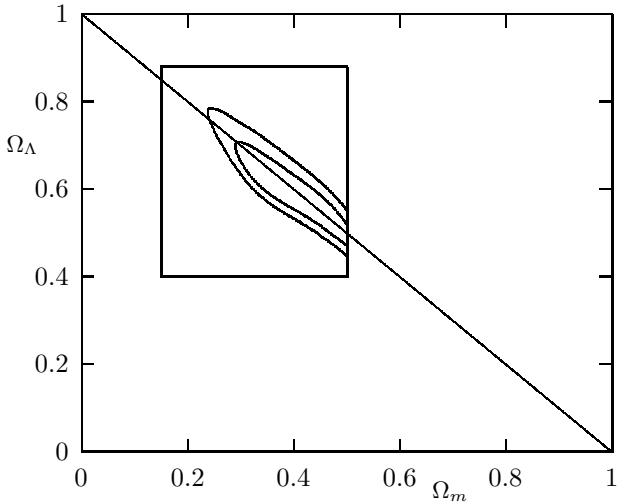

\input 4cons12.tex
\caption{The confidence contours of the log-likelihood sum of
MAXIMA-1, early CMB, Double Radio Galaxies,
and LSS (including BOOMERANG and various other independent
constraints). The curves correspond to 1$\sigma$ and 2$\sigma$ in
the ($\Omega_m, \Omega_{\Lambda}$)-plane. The significance of the
square is described in the text. The diagonal line corresponds to
a flat cosmology.} \label{fig-2}
\end{figure}

In Fig. \ref{fig-2} we show the confidence contours of the log-likelihood
sum of the data sets discussed in this subsection and in the CMB
subsection. We have plotted our polynomial approximation only in
the ranges of $\Omm$ and $\Oml$ that we sample. As can be clearly
seen, the likelihood function contains information mainly on
$\Om0$, but it also gives a rather conspicuous upper limit on the
orthogonal combination $\Oml - \Omm$.

\section{RESULTS}

\begin{figure}[t]
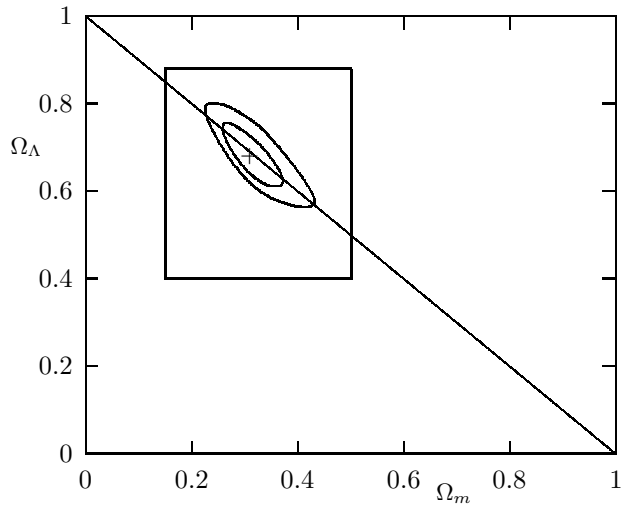

\input mlfit.tex
\caption{Figs. \ref{fig-1} and \ref{fig-2} combined. The '+' marks the best fit:
($\Omega_m, \Omega_{\Lambda}$) = (0.31,0.68). The diagonal line
corresponds to a flat cosmology.} \label{fig-3}
\end{figure}

Adding up our polynomial approximations to the confidence
contours of all the data discussed in the previous section,
results in Fig. \ref{fig-3}, where we show the location of the minimum, the
$1\sigma$ and $2\sigma$ contours. From this Figure one can read
off the following results:
\begin{eqnarray}
\Omm = 0.31 ^{+0.12}_{-0.09}\\
\Oml = 0.68 \pm 0.12 ,
\label{f2}\end{eqnarray}
or alternatively
\begin{eqnarray}
\Om0 = 0.99 \pm 0.04 \\
\Oml - \Omm = 0.37 ^{+0.20}_{-0.23} .\label{f3}\end{eqnarray}

Of these results, only the determination of $\Om0$ is quite
precise and worth detailed attention. We can conclude from it
that a flat universe with $\Om0 = 1$ is very likely.

To the results in Fig.\ref{fig-3} we have to add some further
quantifiable systematic errors which we evaluate as follows.

As mentioned earlier, Perlmutter et al. (1998, 1999) have quoted a
total systematic error for $\Omm^{flat}$ and $\Oml^{flat}$ along
the flat line of $\pm 0.05$. We consider that the same error
should be applied to the SN Ia data of HSST, where a similar
evaluation did not give a significant result due to the limited
sample of SNe Ia. As explained in Sec. 3.1, this is further
motivated by the discord between the MLCS method and the template
method. Displacing both the SN Ia contours by $\pm 0.05$ along
the flat line, we obtain a very small systematic error in the
$\Om0$ direction
\begin{eqnarray}
\Delta_1(\Om0) =\  ^{+0.012}_{-0.006} . \label{f4}\end{eqnarray}

There are also two kinds of systematic errors inherent to our
method of analysis. Firstly, we are reading off the coordinates
of the confidence contours of the different observations with some
finite precision. We estimate this precision to be
\begin{eqnarray}
\Delta_2(\Om0) = 0.027 . \label{f5}\end{eqnarray}

Secondly, since we only use 20 points to fit the confidence
contours of each observations, there is an arbitrariness in their
choice; all we require is that the confidence contours should be
well fitted by whichever polynomial. We have tested this
polynomial arbitrariness and found that it results in the
systematic error
\begin{eqnarray}
\Delta_3(\Om0) = 0.01 . \label{f6}\end{eqnarray}

The quadratic sum of the errors in Eqs. (\ref{f4}), (\ref{f5}),
(\ref{f6}) is then
\begin{eqnarray}
\Delta_{tot}(\Om0) = 0.03 . \label{f7}\end{eqnarray}

Thus our final result for $\Om0$ is
\begin{eqnarray}
\Om0 = 0.99 \pm 0.04 \pm 0.03 , \label{f8}\end{eqnarray} where
the first error is statistical and the second error systematical.
Thus our total error is $\pm 0.05$.

Let us now turn to the case of exact flatness, $\Omm = 1 - \Oml$.
 Along the flat line the SN Ia systematic error is
\begin{eqnarray}
\Delta_1(\Om0)^{flat} = \pm 0.025 . \label{f9}\end{eqnarray}

Our result is then
\begin{eqnarray}
\Omm^{flat} = 0.31 \pm 0.04 \pm 0.04, \label{f10}\end{eqnarray}
where the first error is statistical and the second error
systematical. Thus our total error here is $\pm 0.055$. Note
once again that this systematic error is not included in Fig.\ref{fig-3}.

\section{DISCUSSION}

The closest comparison we can make with other analyses of $\Om0$
is that of Jaffe et al.(2000), who combine BOOMERANG, MAXIMA-1,
COBE DMR, the SN Ia constraints of SCP (but not HSST), and another
LSS input than we have used. Also, the earlier CMB data
compiled by Lineweaver (1998) and Tegmark (1999), and the
information from the double radio sources are not used. They find
\begin{eqnarray}
\Om0 = 1.06 \pm 0.04 . \label{f11}\end{eqnarray} to be compared
with our result in Eq. (\ref{f8}). The difference in central
value is easy to understand, and due to two causes. The main
cause is the way BOOMERANG and MAXIMA-1 have been combined, as we
explained already in the Data section. A small shift in the same
direction is due to our inclusion of other input.The
statistical errors are the same, but we have in addition a
systematic error, part of which is not applicable to the Jaffe et
al.(2000) analysis.

A rather similar analysis is that of Durrer \& Novosyadlyj
(2000), who find $\Omm + \Oml \approx 1.06$ and $\Omega_k \approx
-0.06$, where our definition of $\Om0$ corresponds to $\Omm +
\Oml + \Omega_k = 1$. On the flat line these authors find $\Omm =
0.35 \pm 0.05$. We also note that these authors have taken into
account all the observational constraints we used in our previous
analyses (Roos \& Harun-or-Rashid 1998, 1999, 2000), and which we
therefore did not refer to explicitly here.

One further constraint which we did not make use of here is the
position of the peak in the matter power spectrum of quasars at
$z\approx 2$ as observed by Roukema \& Mamon (2000). The reason
for the omission is that their likelihood contours in the $\Omm ,
\Oml$ space are so jagged that our polynomial approximations just
cannot reproduce them. To quote one result, they find
\begin{eqnarray}
\Omm^{flat} = 0.30 \pm 0.15 \end{eqnarray} in good agreement with
us. The error here is so large that the inclusion of this result
would not have changed our conclusions.

As we mentioned in Sec. 3.1, Helbig (2000) (fig. 3) has plotted
preliminary lensing constraints from the Cosmic Lens All-Sky
Survey (CLASS) of radio lenses, which appear to be in strong
conflict with the SNe Ia data. A joint analysis is statistically
quite meaningless, but excluding one or the other is a biased
choice. Since we have so far included the SNe Ia data and
excluded the lensing data, let us now include both. The result
is shown in Fig. \ref{fig-3a}. One notes that the best value then moves to
$\Omm = 0.34 ^{+0.11}_{-0.03}, \ \Oml = 0.63 ^{+0.04}_{-0.10},$ or
alternatively $\Om0 = 0.97  ^{+0.05}_{-0.04}, \
\Oml - \Omm = 0.29 ^{+0.05}_{-0.18}.$
We note that this best value is excluded by the SNe Ia data
at 1$\sigma$ CL, and it is excluded by the lensing data at 97\% CL.

We conclude from Sec. 4 that $\Om0$ is equal to unity to
within $\pm 0.05$ and that $\Omm^{flat} = 0.31$ to within $\pm
0.055$.

\label{conc}

\begin{acknowledgements}The authors wish to acknowledge helpful
correspondence with B. Novosyadlyj and S. Weinberg. We wish to thank
the referee for his useful suggestions. S. M. H. is
indebted to the Magnus Ehrnrooth Foundation for support.

\end{acknowledgements}

\begin{table}
\begin{center}
\begin{tabular}{lll}
\hline {\em Observation} & {\em Reference}  & {\em Source}\\
\hline
SN Ia: HSST       & Riess et al. (1998)  & Fig. 6    \\
SN Ia: SCP        & Perlmutter et al. (1999) & Fig. 7 \\
CMB: MAXIMA-1    & Balbi et al. (2000) & Fig. 3  \\
CMB compilation    & Tegmark (1999)  & Fig. 3   \\
Double Radio Gal.    & Guerra et al. (2000)  & Fig. 11   \\
LSS         & Novosyadlyj et al. (2000b) & Fig. 1(a) \\
LENSING     & Helbig (2000)  & Fig. 3  \\
\hline
\end{tabular}
\end{center}
\caption {The observational data used in the fifth order polynomials
are summarized. }
\label{table1}
\end{table}

\begin{figure}[t]
\input withhel.tex
\caption{LENSING (Helbig 2000) is combined with fig. \ref{fig-3} The '+' marks
the best fit: ($\Omega_m, \Omega_{\Lambda}$) = (0.34,0.63). The diagonal
line corresponds to a flat cosmology.} \label{fig-3a}
\end{figure}

\end{document}